# Phase-slip induced dissipation in an atomic Bose-Hubbard system

D. McKay[1], M. White[1], M. Pasienski[1] & B. DeMarco[1]

[1]*Department of Physics, University of Illinois at Urbana-Champaign, 1110 W Green St, Urbana, IL 61801*

**Phase slips play a primary role in dissipation across a wide spectrum of bosonic systems, from determining the critical velocity of superfluid helium[1] to generating resistance in thin superconducting wires[2]. This subject has also inspired much technological interest, largely motivated by applications involving nanoscale superconducting circuit elements, e.g., standards based on quantum phase-slip junctions[3]. While phase slips caused by thermal fluctuations at high temperatures are well understood[4], controversy remains over the role of phase slips in small-scale superconductors[5]. In solids, problems such as uncontrolled noise sources and disorder complicate the study and application of phase slips[6]. Here we show that phase slips can lead to dissipation for a clean and well-characterized Bose-Hubbard (BH) system by experimentally studying transport using ultra-cold atoms trapped in an optical lattice. In contrast to previous work, we explore a low velocity regime described by the 3D BH model which is not affected by instabilities, and we measure the effect of temperature on the dissipation strength. We show that the damping rate of atomic motion—the analogue of electrical resistance in a solid—in the confining parabolic potential fits well to a model that includes finite damping at zero temperature. The low-temperature behaviour is consistent with the theory of quantum tunnelling of phase slips, while at higher temperatures a cross-over consistent with the transition to thermal activation of phase slips is evident. Motion-induced features reminiscent of vortices and vortex rings associated with phase slips are also observed in time-of-flight imaging.**

**These results clarify the role of phase slips in superfluid systems. They may also be of interest to outstanding questions regarding dissipation in other bosonic systems, such as the source of metallic phases observed in thin films[7,8], and might serve as a test bed for theories of bosonic dissipation based upon variants of the BH model[9].**

Although believed by many to be the simplest model that captures the relevant features of boson physics in a variety of physical systems, the BH model is not integrable, and therefore a full characterization of its features is a challenging theoretical problem. The BH model is described by the Hamiltonian

$$H = -J\sum_{\langle ij \rangle}\left(b_i^\dagger b_j + b_i b_j^\dagger\right) + \sum_i \epsilon_i n_i + U/2\sum_i n_i(n_i - 1), \quad (1)$$

where $b_i$ removes an atom from site $i$, $n_i = b_i^\dagger b_i$ is the number of bosons on site $i$, and $\varepsilon_i$ is the energy cost for a boson to occupy site $i$. Particles in the BH model move by tunnelling with energy $J$ between adjacent lattice sites $i$ and $j$ and interact pairwise on the same site with energy $U$. In our experiment we can directly test this model because ultra-cold atoms trapped in an optical lattice are a realization of the BH model (for sufficient lattice potential depth) [10,11]. While material and electronic parameters are not easy to independently configure in solid systems, we are able to quantitatively probe the BH model by precisely determining and controlling parameters such as $J$ and $U$.

We study mass transport, the equivalent of charge transport for neutral bosons. Charge transport is studied in solids by using an electric field to apply a uniform force to the charge carriers. In contrast, we use the spatially inhomogeneous restoring force from a parabolic confining potential to excite damped harmonic motion of the atom gas center-of-mass (COM). The confining potential is included in Eq. 1 through the site energies $\varepsilon_i = \frac{1}{2}kr_i^2$, where $r_i$ is the distance from site $i$ to the centre of the harmonic potential and *k* is the spring constant for the restoring force. We measure the COM motion damping rate for small COM velocities. In this regime—analogous to the linear, or ohmic, regime for conductivity experiments on solids—the damping rate is independent of velocity. In analogy to experiments on solids, we measure the response of the damping rate to changes in the temperature of the atom gas. We also extend measurements possible in solids by examining how the damping rate depends on the ratio *J*/*U*, controlled by tuning the lattice potential depth.

Previous work on transport of BECs in optical lattices has focused on regimes not described by the BH model[12], on low-dimensional systems[13,14], on relatively high-velocity transport[15], and on probing the Landau and dynamic instabilities[16-19]. In contrast to much of that work, we work at low velocity and do not study phenomena associated with either instability. The maximum COM velocity in our data is controlled to be smaller than the critical velocity for the Landau and dynamic instabilities[20], including the effect of strong interactions[21]. We do not observe any phenomena characteristic of these instabilities, such as significant change in the condensate fraction, strong non-linear damping, or excitations of the condensate similar to those observed in ref. 17. Measurements of temperature-dependent dissipation in solids have proven to be a powerful tool in understanding phenomena such as phase slips[5] and other sources of resistance[7]; our work is the first systematic investigation of the effect of temperature on transport in an optical lattice.

The experimental sequence is shown in Fig. 1. A $^{87}$Rb Bose-Einstein condensate (BEC) is created in a Ioffe-Pritchard magnetic trap using standard cooling and trapping techniques (see ref. 22 for a description of our BEC apparatus). The BEC is adiabatically transferred into a cubic optical lattice with *d*=405.9 nm between sites; three atoms or less are confined on each lattice site for the data in this paper. The strength of the lattice potential is characterized by a dimensionless parameter *s*, defined by the lattice potential depth $sE_R$ along each lattice direction ($E_R = \hbar^2 \pi^2 / 2md^2$ is the recoil energy and $m$ is the atomic mass). After transfer into the lattice, COM motion is generated by applying a rapid impulse to the BEC along the vertical direction $z$. The COM velocity is measured using time-of-flight (TOF) imaging after the motion is allowed to freely evolve for up to 200 ms.

The COM motion we observe is described well by damped harmonic motion; Fig. 2 shows representative data. We fit the time evolution of the BEC COM velocity assuming the equation of motion $m^* \ddot{z} = -2m^* \gamma \dot{z} - kz$ for the COM coordinate $z$ in the impulse direction, where $m^*$ is the effective mass. The damping rate $\gamma$, the oscillation frequency $\omega_d = \sqrt{k/m^* - \gamma^2}$, and the initial velocity (which ranges from ~0.8–1.8 mm/sec for the data in this paper) are left as free parameters in the fit. Our model of the COM motion assumes that the restoring force $-kz$ from the harmonic confinement and a dissipative force $-2m^* \gamma \dot{z}$ act on the BEC COM. The damping rate $\gamma$, which is the exponential decay rate for the amplitude of the motion, is the equivalent of electrical resistance in a solid. This can be understood by considering that Ohmic resistance in a material—regardless of the source of dissipation—leads to a force on the charge carriers proportional to $-\rho v$, where $\rho$ is the resistivity and $v$ is the charge carrier velocity.

The temperature dependence of the damping rate $\gamma$ is shown in Fig 3. We measure the ratio of temperature $T$ to the critical temperature $T_c$ for condensation in the magnetic trap by determining the fraction of atoms in the BEC before transfer into the lattice. In Fig. 3 we show the measured $\gamma$ for $s=2$ and 6, lattice depths which sample only the superfluid region of the BH phase diagram at zero temperature[9]. Equation 1 requires beyond tight-binding tunnelling energy corrections for $s=2$ that must be considered for detailed comparison between theory and the data shown in Fig. 3(b). The data are shown vs. inverse temperature because we fit the data to a model, $\gamma = \gamma_0 + [\gamma_\infty - \gamma_0] e^{-\Delta E / (T/T_c)}$, of thermally activated damping which permits a finite zero-temperature damping rate $\gamma_0$. The data show excellent agreement with this model, excluding the highest temperature data point for each *s*. The fitted $\gamma_0$, 1.4±0.4 Hz for *s*=2 and 13±2 Hz for *s*=6, are inconsistent with zero at greater than a 95% confidence level. Consistency with a model that includes finite dissipation at zero temperature is, by definition, the equivalent of metallic behaviour in a solid—metals, superconductors, and insulators are defined as materials that possess finite, zero, and infinite resistance at zero temperature, respectively[7]. We have verified that the levelling off of $\gamma$ at low temperature is not caused by $T/T_c$ saturating in the lattice by measuring BEC fraction after release from the lattice. The BEC fraction does not change significantly after transfer into the lattice for the range of *s* and $T/T_c$ in Fig. 3.

The data shown in Fig. 3 are qualitatively consistent with a predicted cross-over between quantum tunnelling and thermal activation of phase slips[21]. Phase-slip models of dissipation were first used to describe the intrinsic critical velocity of superfluid $^4$He [1] and the appearance of resistance in a superconducting wire[2]. In the context of this work, phase slips permit the BEC COM velocity to relax through generation of topological phase structures such as vortices and vortex rings. The COM motion of the BEC is metastable because there is an activation barrier to a phase slip event occurring and driving the system toward the zero velocity ground state. Phase slips occur when the system tunnels through the activation barrier or when thermal fluctuations produce the required activation energy. The dissipation rate should therefore cross-over from thermally activated behaviour (i.e., exponentially dependent on the inverse temperature) to temperature independent at a characteristic temperature for any system in which phase slips are the dominant dissipation mechanism. The data in Fig. 3 show excellent agreement with this cross-over behaviour, except at the highest temperature.

Other dissipation mechanisms may contribute to the damping rate at high temperature. For example, mutual friction between the BEC and thermal component may play a role at high temperature and large *s*, where we observe relative motion between the BEC and thermal component (see Fig. 2b). Mutual friction, first observed in second-sound experiments in superfluid $^4$He [23] and later in experiments with atomic BECs[24,25], arises from interactions between atoms in the BEC and the thermal component. Because we observe the thermal component typically immobilized at high temperature and large *s* (similar behaviour was observed in ref. 26), mutual friction will tend to bring the BEC to rest. Mutual friction, or any dissipation mechanism induced by the thermal component, cannot explain the measured $\gamma_0$, since the thermal component vanishes at zero temperature.

In the quantum tunnelling regime, the phase slip rate is predicted to be proportional to $e^{-S}$, where *S* is an action characterizing the process of quantum fluctuations driving the system to lower velocity[2,27]. A generic scaling law for the action, $S \propto \sqrt{J/U}$, is derived for the BH model in the appendix to ref. 21. In Fig. 4 we plot the measured damping rate for different $S \propto \sqrt{J/U}$ at two temperatures in the temperature-independent damping regime. The data are fit to a line on the log-lin scale, which is equivalent to the model $\ln \gamma \propto \sqrt{J/U}$ predicted for quantum tunnelling of phase slips; the data show excellent agreement with the predicted scaling law. The systematic increase in the damping rate for the higher temperature data in Fig. 4 is consistent with a residual rate of thermally activated phase slips.

The insets to Fig. 4 show features consistent with a vortex ring and a vortex directly observed in BECs undergoing COM motion. The cores of vortex rings and vortices are visible in TOF images if the free expansion time is long enough for the expanded core size to be optically resolved and if the core aligns properly with the imaging axis[28,29]. The insets to Fig. 4 show typical TOF images in which vortex rings or vortices are observed and are taken after 50 ms of free expansion following 2 ms of COM motion evolution at *s*=8. The features observed in these images consistent with a vortex ring viewed on edge and a single vortex observed along its core are marked in red and were not detectable at the shorter expansion time used for the data in Fig. 3 and Fig. 4. Vortex rings and single vortices nucleated by phase slips must lie in a plane perpendicular to the direction of COM motion; single vortex lines may be oriented along any direction in this plane. Vortex rings will therefore be detected on edge by our imaging system, but single vortices will only rarely align with the imaging axis and be clearly resolved. We observe these features with approximately 20% probability at $s$=8, consistent with a random generation process such as phase slips; we do not observe vortex features if COM motion is not excited. We find that vortices and vortex rings are most likely to be detected close to the edge of the BEC where the energy gap to nucleation is smallest.

In conclusion, we observe temperature-dependent damping of COM motion in an optical lattice that is consistent with dissipation caused by phase slips. The parabolic potential used to confine the atoms gives rise to an inhomogeneous density distribution, which may enhance the effect of phase slips because the activation barrier is suppressed at the edge of the BEC. This system may therefore be comparable to thin superconducting wires and strips, in which vortices entering and leaving at the boundaries strongly influence current flow[30]. The effects of inhomogeneous density and finite size on phase slip dynamics in this system remain to be conclusively addressed theoretically. The technique used in this paper can be extended to probe transport properties in models that are relevant to solid materials, such as two-dimensional and disordered BH models. Direct imaging of vortex rings and vortices nucleated by phase slips may be used to address the microscopic dissipation dynamics in this system.

**Methods Summary**

The optical lattice is created using three pairs of orthogonally polarized laser beams at 812 nm. These beams are weakly focused to a 120 μm waist, and slightly frequency offset to eliminate residual cross-dimensional interference resulting from imperfect polarization. COM motion is excited by changing the strength of a confining magnetic potential for 5 ms. The signal-to-noise ratio for measuring the COM velocity is improved by turning off the lattice potential in 200 μs, thereby eliminating atoms effectively diffracting from the lattice potential.

Temperature is determined by measuring the fraction of atoms in the BEC. For our lowest temperature data ($T/T_c$=0.6 and 0.3), accurately detecting the atoms in the thermal component requires images taken at high and low optical depth.

The oscillation data used to measure $\gamma$ were checked in two ways for non-linear response. First, no significant change in the fitted $\gamma$ was measured if the first period of motion was excluded from the fit. Secondly, fitting the data to a non-linear damping model with $\gamma = \Gamma |\dot{z}|^\alpha$ typically increased reduced $\chi^2$ for the fit by 10–20%. We found no clear dependence of $\alpha$ on $T/T_c$ or $J/U$, and the fitted value of $\alpha$ averaged across all of the data was 0.68±0.07. Ultimately, our sensitivity to weak velocity dependence is limited because of finite signal-to-noise ratio in measurements of the COM velocity.

**Acknowledgements** We thank R. Barankov, E. Demler, P. Goldbart, N. Goldenfeld, D. Pekker, and P. Phillips for stimulating discussions and D. S. Jin, N. Mason, and J.V. Porto for critically reading this manuscript. This work was supported by the National Science Foundation, the Office of Naval Research, and the UIUC Research Board. D. McKay acknowledges support from the Carver Foundation and NSERC.

**Author Information** Correspondence and requests for materials should be addressed to B. DeMarco (bdemarco@uiuc.edu).

**Figure 1. Schematic of experimental apparatus and experimental sequence .** A $^{87}$Rb BEC (blue sphere) is transferred over 100 ms from a magnetic trap (copper-coloured) into an optical lattice by superimposing three pairs of counter-propagating lattice beams (red). A 7 ms impulse is applied to excite COM motion in the overall potential *V* experienced by the atoms, which is the combination of the lattice potential with a parabolic confining potential provided by the magnetic and optical dipole trap. After allowing up to 200 ms for the motion of the BEC to evolve, the lattice and magnetic trap are turned off and the BEC is released. The COM velocity of the BEC is measured after 24 ms of free expansion by taking an image. The inset is not to scale—the BEC diameter ranged from 30–40 μm for the data in this paper (corresponding to 75–

100 lattice sites) and the maximum amplitude of COM motion in the harmonic potential was ~10 μm.

**Figure 2. Representative data used to measure the damping rate $\gamma$.** Each data point represents the result of one experimental cycle, in which the BEC (solid circles) and thermal component (hollow circles) COM velocity are measured. **a,** Data for $s=2$, $T/T_c=0.85$, and $8\times10^5$ BEC atoms and **b,** for $s=6$, $T/T_c=0.93$, and $2.7\times10^5$ BEC atoms. The error bar in **a** is the standard deviation in the COM velocity if no impulse is applied. The solid line is a fit to the data used to determine the damping rate $\gamma$. False colour images that are used to measure velocity are shown as insets to **b**; the dashed lines bisecting the 0.47 mm field-of-view are a guide to the eye. Regions of high density are coloured red, while low density regions are blue. The narrow (broad) component in each image is the BEC (thermal component).

**Figure 3. Temperature dependence of the damping rate $\gamma$ for a, $s=6$ and b, $s=2$.** Each data point represents the value of $\gamma$ determined from BEC COM motion at a fixed temperature and $s$. The red data points are excluded from the fit (solid line) to a model of thermally-activated damping. The fitted energy gap $\Delta E$ is $k_B\times0.5\pm0.1$ μK and $k_B\times0.33\pm0.08$ μK for $s=2$ and $s=6$, respectively ($k_B$ is Boltzmann's constant). The error bars in $\gamma$ represent the uncertainty in the fit to COM velocity data. The standard deviation in the data used to determine $T/T_c$ for each data point are used for the error bars in $(T/T_c)^{-1}$.

**Figure 4. Scaling of $\gamma$ with $\sqrt{J/U}$.** The tight-binding limit is used to calculate $\sqrt{J/U}$ from $s$. The solid points are taken with $T/T_c=0.76$ and $3\times10^5$ BEC atoms and the hollow points with $T/T_c=0.312$ and $1.1\times10^6$ BEC atoms. The error bars in $\gamma$ are the uncertainty from the fit to the COM velocity; there is less than 3% uncertainty in $\sqrt{J/U}$. The -1.6±0.1 slope determined by the linear fit (solid line)

on the log-lin scale is consistent between the two temperatures. TOF images in which features consistent with a vortex ring and single vortex are observed are shown in false colour in the insets **a** and **b**, respectively; other excitations present in **b** are not clearly resolved. The images are fit to a Thomas-Fermi profile, and the residual is shown in greyscale. The edge of the BEC determined from the fit is shown as a white ellipse in each inset.

**Methods**

**Optical lattice apparatus.** The details of the apparatus used to create optical lattices in our experiment have not been published elsewhere. $^{87}$Rb BECs are created in the $|F=1, m_F=-1\rangle$ state in a magnetic trap with a 21.7 Hz radial frequency and 14.5 Hz axial frequency. The optical lattice potential, $-s\left[\cos^2(\pi \hat{n}_1 \cdot \vec{r}/d) + \cos^2(\pi \hat{n}_2 \cdot \vec{r}/d) + \cos^2(\pi \hat{n}_3 \cdot \vec{r}/d)\right]$, is formed from three retro-reflected laser beams with orthogonal wavevectors and polarizations along directions $\hat{n}_{1,2,3}$. The beam directions make 60° and 45° angles with the vertical direction *z*. Two of the optical lattice beams are frequency offset by +19.66 and -176.94 MHz from the third to prevent residual cross-dimensional interference. The magnetic trap is left on during and after transfer into the lattice to support the atoms against gravity. The impulse used to excite COM motion is created by briefly changing the equilibrium position of the BEC in the magnetic trap. The harmonic potential is not spherically symmetric as we specified for simplicity in Eq. 1. The fractional difference between the weakest and the geometric mean of the oscillator frequencies along the lattice directions is 8% for *s*=2, decreasing to 2% by *s*=6. Eq. 1 also requires corrections for next-nearest-neighbour tunnelling for *s*=2.

**COM velocity measurement.** Standard absorption imaging is used to measure the COM position of the atoms, which is translated into COM velocity, after release from the optical lattice and harmonic potential. Before turning off the magnetic trap, the lattice is linearly ramped off in 200 μs to improve the image signal-to-noise ratio. We checked that this "band-mapping"[31] step does not affect the measured COM velocity by comparing to data for which the lattice potential was turned off in 350 ns. Images are fit to a Thomas-Fermi (TF) profile if only a BEC is visible, or the combination of a TF and a Gaussian profile if atoms in a thermal component are present. While the COM of each component is left as a free parameter in the fit, we do not analyze motion of the thermal component in this paper.

**Temperature measurement.** We determine temperature for $T/T_c$>0.6 by measuring the fraction of atoms in the BEC using a single absorption image taken after release from the magnetic trap. The atoms are optically pumped (or "repumped") to the $|F=2, m_F=-2\rangle$ state before imaging. The thermal fraction at our lowest temperatures is not visible for images in which the maximum optical depth (OD) is small enough to accurately measure the BEC number. Therefore, the condensate fraction at $T/T_c$=0.312 and $T/T_c$=0.6 is determined by taking two images. One image is partially repumped and used to determine the total OD of the BEC. The second image is fully repumped and is used to determine the total OD of the thermal component by fitting the low-OD region of the image. We compare OD from the two images by calibrating the fractional change in OD for partial repumping. The calibration is performed using a thermal gas and taking images at two free expansion times.

We measure $T/T_c$ using BEC fraction because techniques for determining $T$ in an optical lattice have not been proven. Interactions may change $T/T_c$ for BECs loaded into the lattice for the highest value of $s$ in Fig. 4 [32-34]; the extent to which this effect plays a role in experiments has not been resolved. The temperature of the gas was controlled by altering the evaporative cooling procedure, resulting in varying BEC number and $T_c$ at fixed $s$. The average $T_c$ for the data in this paper is 0.13 μK, determined from the total number of atoms and the magnetic trap oscillator frequency; $T_c$ spans a 0.07 μK range for the data in Fig. 3. The heating caused by the dissipation observed in our experiment cannot be detected within our experimental uncertainty in condensate fraction or absolute temperature.

**Technical noise.** We rule out several technical noise sources as dissipation mechanisms that could explain damping of COM motion. Anharmonicity in the dipole potential may effectively damp COM motion for large *s*. To check for anharmonic behaviour, we measure COM motion when the retro-reflected lattice laser beams are removed, which eliminates the lattice potential and reduces the depth of the dipole potential by a factor of ~2. Using this technique we measure a $\gamma$ consistent with zero for lattice laser intensities corresponding to *s*=9 (*s*=18 if the retro-reflected beams are present), eliminating trap anharmonicity as an effective dissipation source for the data in Fig. 3. Relative motion between the lattice and harmonic potential or fluctuations in *s* (caused by retro-reflecting mirror motion and lattice laser intensity fluctuations, respectively) can lead to dephasing of dipole mode motion by transferring atoms into states with different $m^*$ in excited bands[35]; we do not, however, observe population outside of the lowest-energy band. The total spontaneous emission rate per atom is less than 0.3 Hz for *s*=6, so momentum diffusion caused by scattering light from the optical lattice laser beams is insignificant. The lattice depth varies by less than 3% across the BEC, so spatial variation in the effective mass can play no role in the dissipation timescales measured in our data.

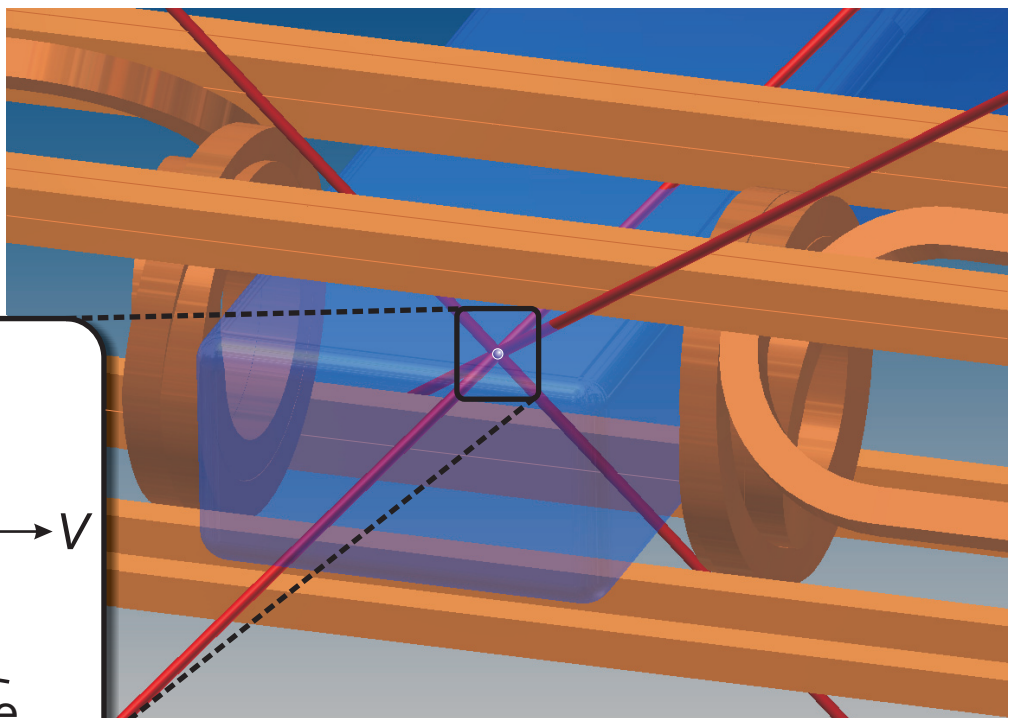
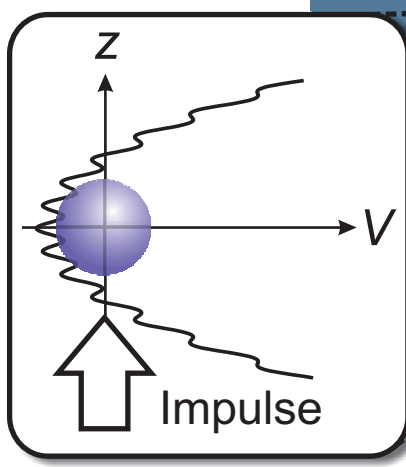
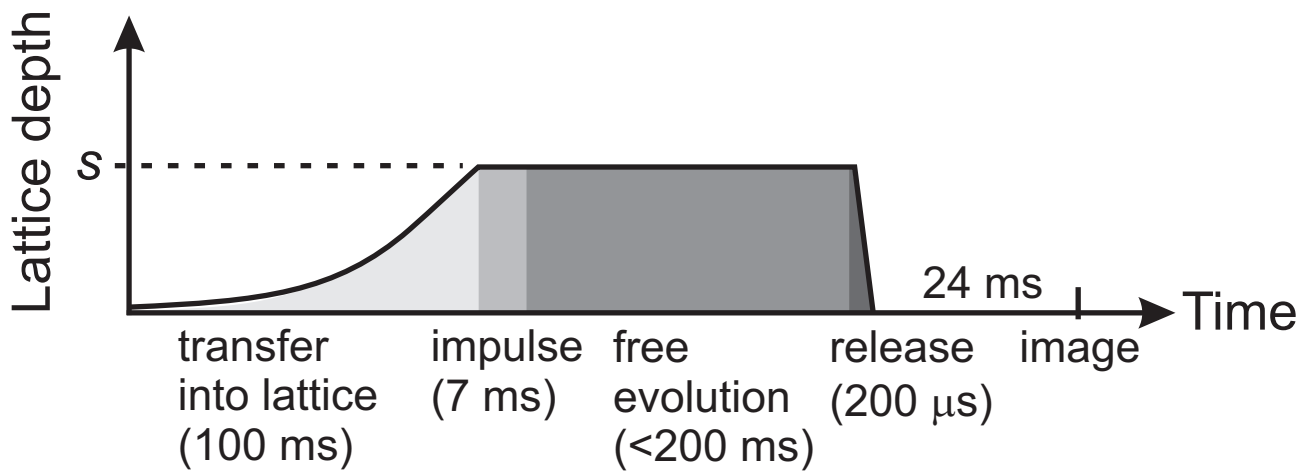

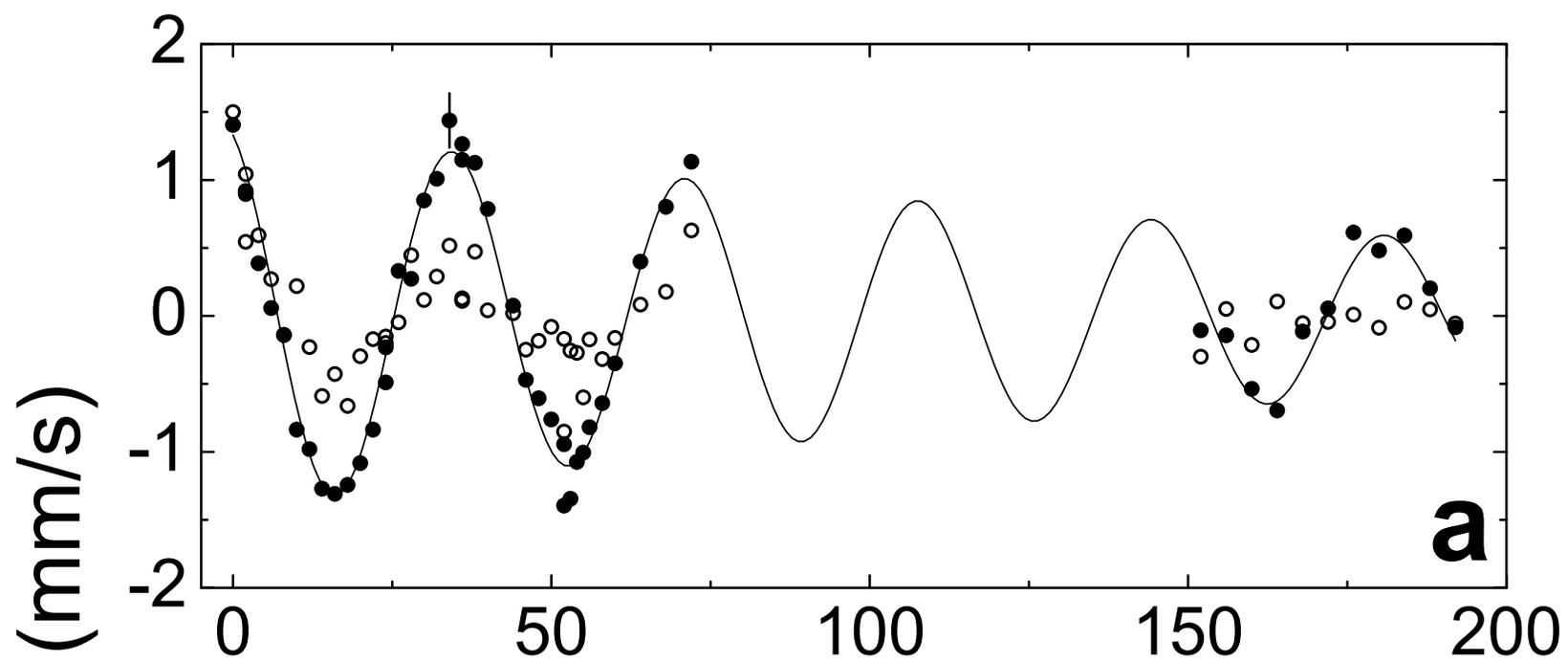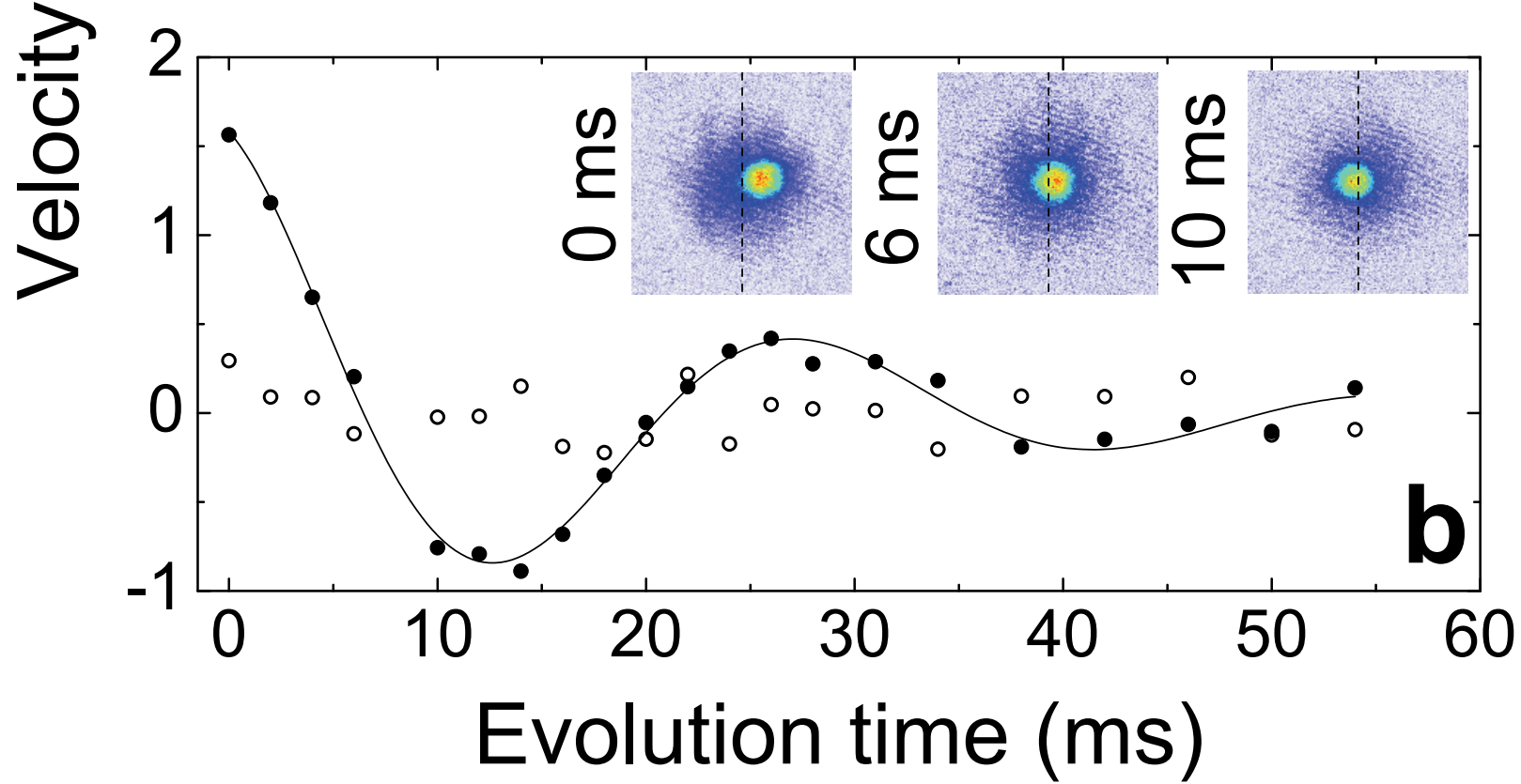

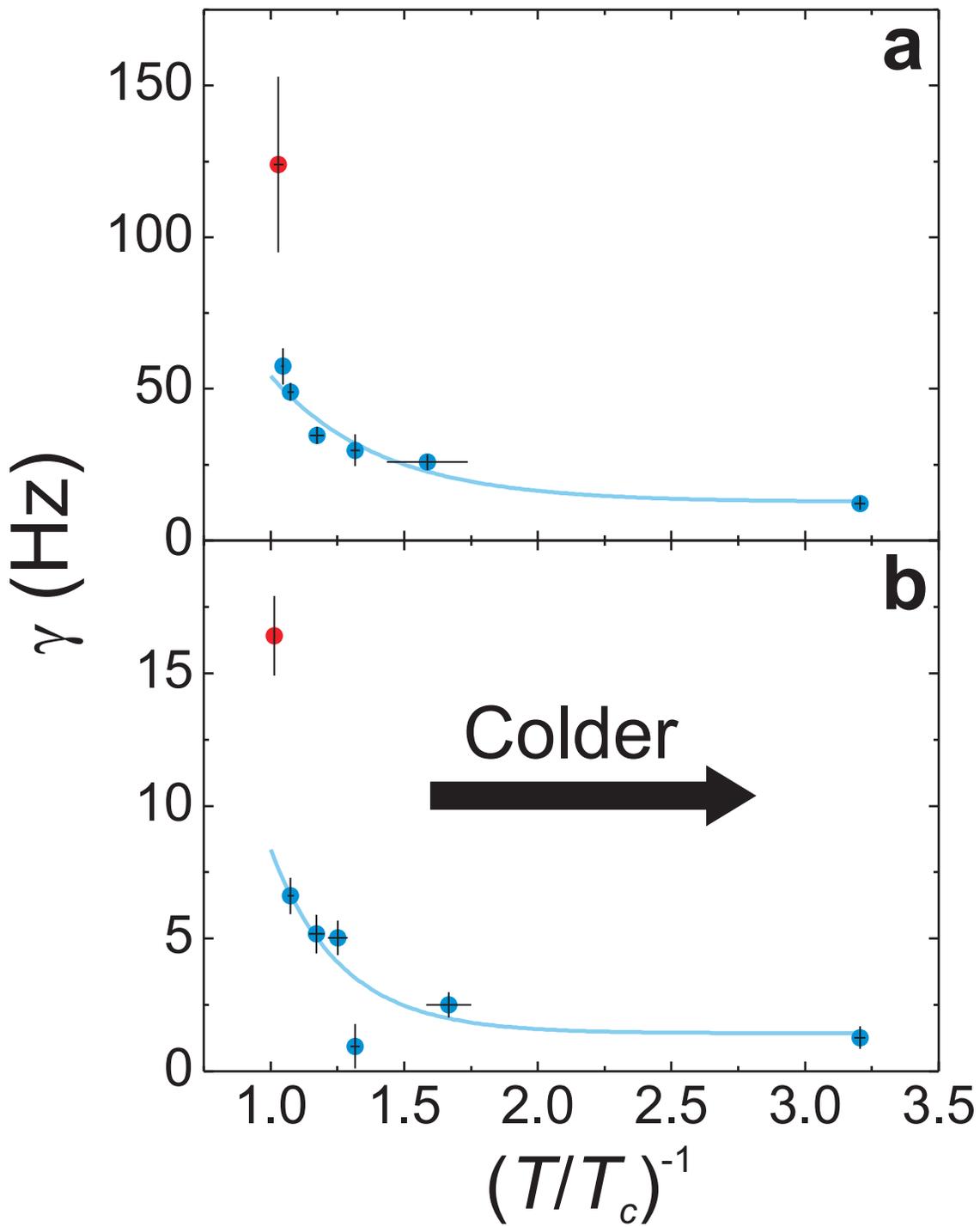

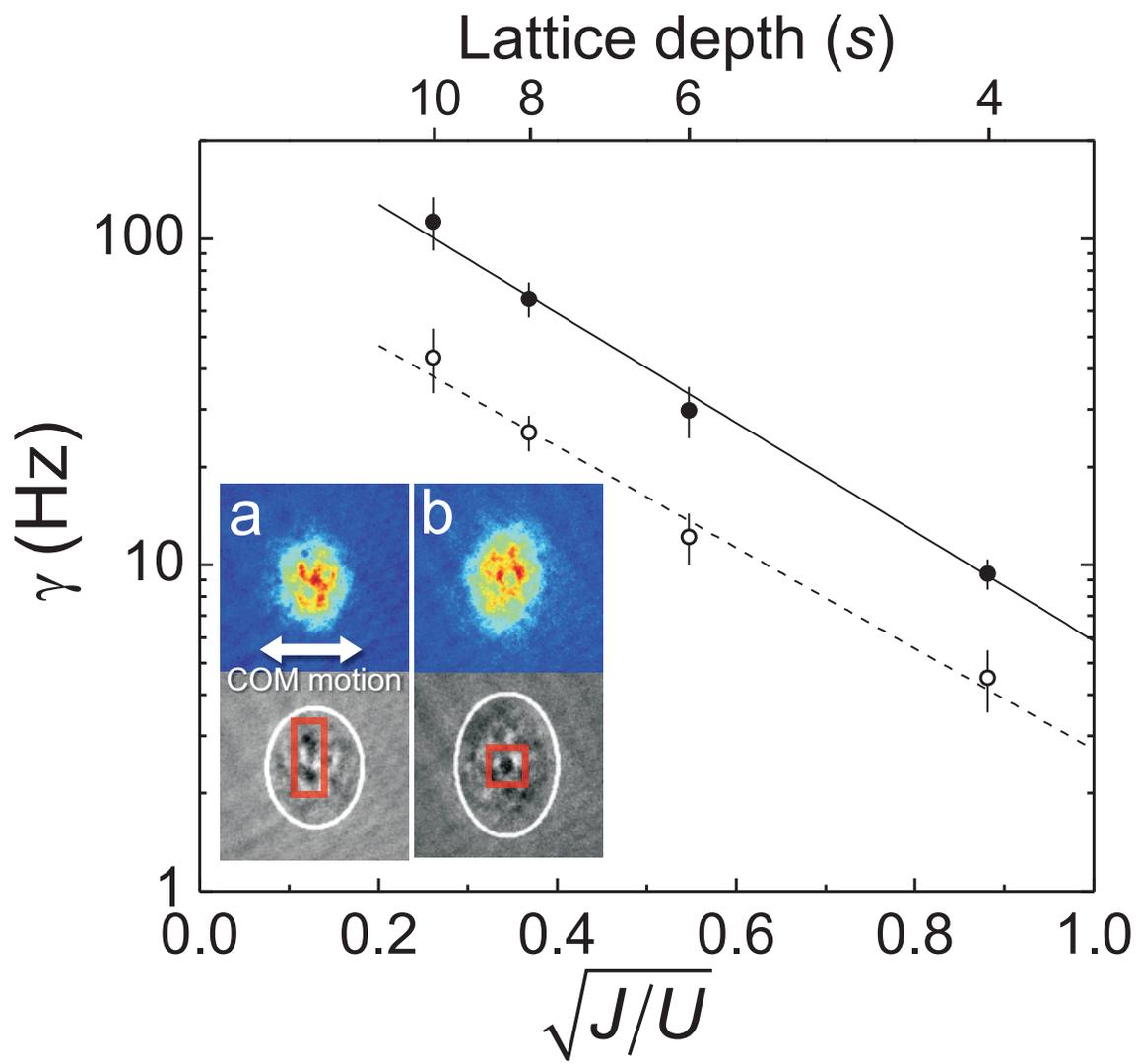